\documentclass{llncs}
\usepackage{times}
\usepackage{url}
\usepackage{bibnames}
 \usepackage{amsmath}
\usepackage{booktabs,multirow,tabularx,tabulary}

\usepackage[]{algorithm2e}
\usepackage{graphics}
\usepackage{wasysym}
\usepackage{units}
\usepackage{slashbox}
\usepackage[title]{appendix}
\usepackage{tkz-euclide}
\usepackage{amssymb}

\begin{document}
\title{Transitivity, Time Consumption, and Quality\\ of Preference Judgments in Crowdsourcing}

\author{Kai Hui\inst{1, 2} \and Klaus Berberich\inst{1, 3}}

\institute{Max Planck Institute for Informatics
\and
Saarbr\"ucken Graduate School of Computer Science
\and
htw saar\\ Saarbr\"ucken, Germany\\
\email{\{khui, kberberi\}@mpi-inf.mpg.de}}

\maketitle

\begin{abstract}
Preference judgments have been demonstrated as a better alternative to
graded judgments to assess the relevance of documents relative to queries. 
    Existing work has verified transitivity among 
preference judgments
  when collected from trained judges,
  which reduced the number of judgments dramatically.
  Moreover, strict preference
  judgments and weak preference judgments, where the latter
  additionally allow judges to state that two documents are equally
  relevant for a given query, are both widely used in literature. 
  However, whether transitivity still holds when collected
  from crowdsourcing, i.e., whether the two kinds of 
  preference judgments behave similarly
  remains unclear.
  In this
  work, we collect judgments from multiple judges using a
  crowdsourcing platform and aggregate them to 
  compare the two kinds of
  preference judgments in terms of transitivity, time consumption, and
  quality. 
  That is, we look into whether aggregated judgments are transitive,
  how long it takes judges to make them, and whether judges agree with
  each other and with judgments from \textsc{Trec}. 
    Our key
  findings are that only strict preference judgments
  are transitive. Meanwhile, weak preference judgments
  behave differently in terms of transitivity, time consumption, as well as of the quality of judgment.
\end{abstract}

\section{Introduction} 
\label{sec.introduction}

\begin{figure}
\centering
\resizebox{.8\textwidth}{!}{
\begin{minipage}[b]{.5\textwidth}
\centering
 \begin{tikzpicture}
    \node (fig) {
      \includegraphics[width=0.24\textwidth]{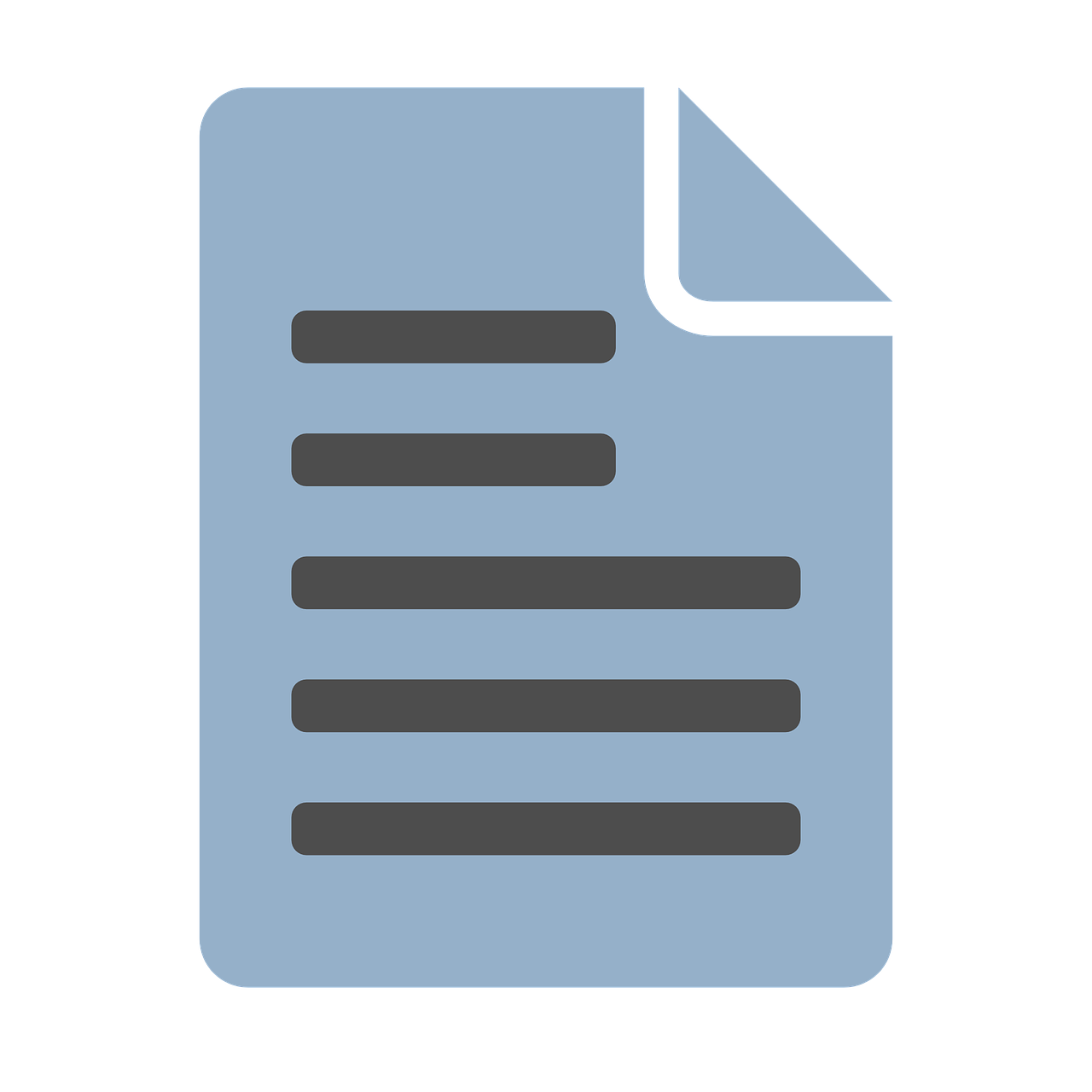}
    };
    \node [anchor=north,color=darkgray,inner sep=0,xshift=0pt,yshift=1pt]
      at (fig.south) {\scriptsize A};
 \end{tikzpicture}

\raggedright
\begin{itemize}
\item[]\textbf{How well does the document A match the query?}
\item[]
\item[$\square$] Highly-Relevant
\item[$\square$] Relevant
\item[$\square$] Non-Relevant
\end{itemize}
\end{minipage}
\begin{minipage}[b]{.5\textwidth}
\begin{minipage}[b]{.4\textwidth}
\centering

 \begin{tikzpicture}
    \node (fig) {
      \includegraphics[width=0.6\textwidth]{figs/singledoc.png}
    };
    \node [anchor=north,color=darkgray,inner sep=0,xshift=0pt,yshift=1pt]
      at (fig.south) {\scriptsize A};
 \end{tikzpicture}
\end{minipage}
\begin{minipage}[b]{.4\textwidth}
\centering
 \begin{tikzpicture}
    \node (fig) {
      \includegraphics[width=0.6\textwidth]{figs/singledoc.png}
    };
    \node [anchor=north,color=darkgray,inner sep=0,xshift=0pt,yshift=1pt]
      at (fig.south) {\scriptsize B};
 \end{tikzpicture}
\end{minipage}
\raggedright
\begin{itemize}
\item[]\textbf{Which document is more relevant  
or they are equivalent to the query?}
\item[]
\item[$\square$] Document A is more relevant
\item[$\square$] Document A and B are equivalent
\item[$\square$] Document B is more relevant
\end{itemize}
\end{minipage}
}
\caption{Examples for graded (left) and preference judgments (right).}\label{figs.jugexemple}

\end{figure}

Offline evaluation in information retrieval following the
Cranfield~\cite{cleverdon1967cranfield} paradigm heavily relies on
manual judgments to evaluate search results returned by competing
systems. The traditional approach to judge the relevance of documents
returned for a query, coined graded judgments, is to consider
each document in isolation and assign a predefined grade (e.g.,
highly-relevant, relevant, or non-relevant) to it. More recently, preference
  judgments have been
demonstrated~\cite{carterette2008here,kazai2013user,Radinsky2011} as a
better alternative. Here, pairs of documents returned for a specific
query are considered, and judges are asked to state their relative
preference. Figure~\ref{figs.jugexemple} illustrates these two
approaches. Initiatives like \textsc{Trec} have typically relied on trained
judges, who tend to provide high-quality judgments. Crowdsourcing
platforms such as Amazon Mechanical
Turk and
CrowdFlower have emerged,
providing a way to reach out to a large crowd of diverse workers for
judgments. While inexpensive and scalable~\cite{alonso2011design},
judgments from those platforms are known to be of mixed
quality~\cite{kazai2011search,moshfeghi2016identify,moshfeghi2016game}.

\newpage

Kazai et al.~\cite{kazai2013user} demonstrated that preference
judgments collected using crowdsourcing can be inexpensive yet
high-quality. In their experiments preference judgments yielded better
quality, getting close to the ones obtained from
trained judges in terms of user satisfaction. 
Unfortunately,
preference judgments are very expensive. 
To judge the relevance of $n$ documents, $\mathcal{O}(n^2)$ preference
judgments are needed, since pairs of documents have to be considered,
whereas $\mathcal{O}(n)$ graded judgments suffice. Luckily, it has
been shown that preference judgments are
transitive~\cite{carterette2008here,rorvig1990simple} when collected
from trained judges, which can be exploited to reduce their required
number to $\mathcal{O}(n\,\log{n})$. Whether transitivity still holds
when preference judgments are collected using crowdsourcing is an open
question as mentioned in~\cite{bashir2013document}. In the aforementioned
studies~\cite{carterette2008here,rorvig1990simple}, trained judges
stated their relative preference for all pairs of documents returned
for a specific query. As a consequence, when considering a triple of
documents, the same judge states relative preferences for all pairs of
documents therein, making transitivity more of a matter of judges'
self-consistency. When using crowdsourcing, in contrast, it is very
unlikely that the same judge states relative preferences for all pairs
of documents from a triple, given that workers typically only
contribute a small fraction of work. Transitivity, if it exists, can
thus only be a result of agreement among different judges. We examine
whether transitivity holds when preference judgments are collected
using crowdsourcing, 
when considering preference judgments aggregated from the stated
preferences of multiple different judges.


Another difference between graded judgments and
preference judgments, as reported by Carterette et
al.~\cite{carterette2008here}, is that preference judgments
tend to be less time consuming. Thus, in their experiments, trained
judges took 40\% less time to make individual preference
  judgments than to make individual graded judgments. We
investigate whether this observation also holds when judgments are
collected using crowdsourcing. If so, there is an opportunity to
reduce cost by paying less for preference judgments.

Beyond that, previous works have considered different variants of preference
judgments. When judges are asked to state strict preferences
for two documents $d_1$ and $d_2$, as done
in~\cite{carterette2008here,Radinsky2011,rorvig1990simple}, they can
only indicate whether $d_1$ is preferred over $d_2$ ($d_1\succ d_2$)
or vice versa ($d_1\prec d_2$). When asking for weak
  preferences, additional options
are provided, allowing judges to state that the two documents are tied
($d_1\sim d_2$)~\cite{kazai2013user,song2011select,zhu2010analysis} 
or two documents are either equally relevant or equally non-relevant~\cite{bashir2013document}. 
Allowing for ties is natural when judging search
relevance, since it is unlikely that each of the possibly hundreds of
returned documents has its own degree of relevance. We investigate
whether weak preferences and strict preferences exhibit transitivity,
and how they compare in terms of time consumption and quality.

\vspace{1mm}


Putting it together, we investigate the following research questions.

\begin{enumerate}
  
\item[] \textbf{RQ1}: \emph{Do weak/strict preference judgments exhibit
  transitivity when collected using crowdsourcing?}\\

\item[] \textbf{RQ2}: \emph{How do weak/strict preference judgments
    compare against graded judgments in terms of time consumption?}\\

\item[] \textbf{RQ3}: \emph{Can weak/strict preference judgments
    collected using crowdsourcing replace judgments by trained
    judges?}\\

\end{enumerate}

\vspace{-5mm}


To answer these, we conduct an empirical study on CrowdFlower. Using
topics and pooled documents from the \textsc{Trec} 
Web Track,\footnote{\url{http://trec.nist.gov/data/webmain.html}} we
collect graded judgments, weak preference judgments, and strict
preference judgments. Akin to Carterette et
al.~\cite{carterette2008here}, we examine transitivity by considering
triples of documents. To analyze the time consumption for different
kinds of judgments, our user interface is carefully instrumented to
record the time that it takes judges to read documents and to make
their judgment. We assess the inter-judge agreement for the different
kinds of judgments and also examine to what extent they can replace
judgments by trained judges from \textsc{Trec}.


We observe that transitivity holds over 90\% for strict preference
judgments collected using crowdsourcing; for weak preference judgments
it only holds for about 75\% of triples. In addition, we find that
judges spend more time when asked for preference judgments than graded
judgments in terms of total time consumption. 
Though time on making a single judgment is
found to be lower for strict preference judgments. 
Finally, we see that
preference judgments collected using crowdsourcing tend to show better
agreement with \textsc{Trec} judges. Moreover, 
the agreement between strict preference judgments from crowdsourcing
and judgments from \textsc{Trec} already match the 
agreement among trained judges reported from literature~\cite{carterette2008here,kazai2013user}.


\textbf{Organization.} The rest of this paper is organized as follows.
Section~\ref{sec.relatedwork} recaps existing literature and puts our
work in context. Following that, in Section~\ref{sec.expdesign}, the
setup of our empirical study is described. Section~\ref{sec.result}
describes its results and provides
answers to the research questions stated above. Finally, in
Section~\ref{sec.conclusion}, we draw conclusions.


\section{Related Work} 
\label{sec.relatedwork}

Preference judgments have been demonstrated as a better
alternative to graded judgments, since there is no need to
define graded levels~\cite{carterette2008here}, their higher
inter-assessor agreement, and better quality~\cite{carterette2008here,kazai2013user,Radinsky2011}.  Moreover,
Carterette et al.~\cite{carterette2008here} pointed out that
preference judgments are less time-consuming than
graded judgments.
\newpage

\textbf{Reduce the number of judgments in preference
    judgments.}  Assuming transitivity can dramatically bring down
the number of judgments from $\mathcal{O}(n^2)$ to
$\mathcal{O}(n\,\log{n})$~\cite{carterette2008here}.  To utilize transitivity,
Rorvig~\cite{rorvig1990simple} verified the transitivity among
judgments from a group of undergraduates.  Carterette et.
al~\cite{carterette2008here} tested transitivity among judgments from
six trained judges, finding that the transitivity holds for 99\% of document triples.  
Different from our settings, both works examined
transitivity with trained judges, which is very different from the
condition under crowdsourcing as indicated in
Section~\ref{sec.introduction}.  Moreover, both works applied
strict preferences in their empirical studies.  Meanwhile,
follow-up works tend to extend this property to weak
  preferences~\cite{song2011select}.  Thus, in this work, we also 
examine the transitivity over weak preference judgments.

\textbf{Weak preferences versus strict preferences.}  The choices
between two kinds of preferences varied a lot among different works,
even though some of them share similar motivations or research
mythologies.  Carterette et al.~\cite{carterette2008here},
Radinsky \& Ailon~\cite{Radinsky2011} and Rorvig~\cite{rorvig1990simple}
employed strict
  preferences in their empirical studies for preference
  judgments.  In the meantime, Kazai et al.~\cite{kazai2013user}
collected weak preference judgments from both trained
judges and crowdsourcing workers to empirically explore the
inter-assessors agreement and the agreement between the collected
judgments and the real user satisfactions.  Song et
al.~\cite{song2011select} introduced an option ``same as'' in the judging
interface and assumed transitivity over the weak preferences
in their \textsc{Quick-Sort-Judge} method.  Additionally, Zhu \&
Carterette~\cite{zhu2010analysis} collected weak preferences
through a ``no preference'' option in their research over the user
preference for the layout of search results.  It seems to us that the
strict and weak preferences are regarded as
interchangeable in existing works. However whether preference
  judgments with and without tie are the same in terms of judgment
quality and judgment efforts remains unclear.

\textbf{Crowdsourcing for relevance judgments.}  Existing works
examined different ways to collect judgments from
crowdsourcing~\cite{grady2010crowdsourcing} and provided a proper
model to follow in collecting graded judgments from
crowdsourcing~\cite{alonso2011design}.  Alonso and
Mizzaro~\cite{alonso2009can,alonso2012using} demonstrated that it is
possible to replace graded judgments from \textsc{Trec} using
crowdsourcing. Additionally, Kazai et al.~\cite{kazai2013user}
compared graded and preference judgments from both
trained judges and crowdsourcing, highlighting that preference
  judgments are especially recommended for crowdsourcing, where
judgment quality can be close to the one from trained judges.
Different from this work, Kaizai et al.~\cite{kazai2013user} measured
agreement based on individual judgments, instead of aggregated 
ones.  As mentioned in~\cite{alonso2012using}, it is the aggregated
judgments that can be used in practice. Moreover, the judgment quality
is measured in terms of the agreement relative to user clicks, whereas
in our work, the measurement is based on judgments from \textsc{Trec}
Web Track. Thereby, in the regards of empirical analysis over judgment
quality, our work can be regarded as an extension to
both~\cite{alonso2012using} and~\cite{kazai2013user}.


\section{Empirical Study on CrowdFlower}\label{sec.expdesign}


\textbf{User interface.} We display queries together with their
description from the \textsc{Trec} Web Track 2013 \&
2014. Judges are instructed to
consider both the query and its corresponding description as in Figure~\ref{figs.jugexemple}. To help
them understanding the topic, we also display a link to run the query
against a commercial web search engine. When collecting preference
judgments, we show the query and description together with two
documents (A and B) and ask judges ``Which document is more relevant
to the query?''. When collecting strict preferences, judges can choose
between the options ``Document A is more relevant'' and ``Document B
is more relevant''. A third option ``Document A and B are equivalent''
is added, when collecting weak preferences. When collecting
graded judgments, the query and description are shown together with a
single document. Judges are asked ``How well does the document match
the query?'' and can click on one of the grades ``Non-Relevant'',
``Relevant'', and ``Highly Relevant''. In our instructions we include the same
definitions of grades from \textsc{Trec}.

\textbf{Quality control.} Unique tasks, in our case judgments, are
referred as rows in CrowdFlower. Multiple rows are grouped into a
page, which is the basic unit for payment and quality control. The
major means to control quality are test questions, that is, rows with
a known expected input from workers. Test questions can be used to
run a qualification quiz, which workers have to complete upfront. By
thresholding on their accuracy in the qualification quiz, unreliable
workers can be filtered out. Moreover, test questions can be
interspersed with rows to continuously control the quality of
work. Workers can thus be banned once their accuracy on interspersed
test questions drops below a threshold. The accuracy threshold is set
as 0.7, following the default on CrowdFlower.

\textbf{Job settings.} When collecting graded judgments a page
consists of eleven judgments and a test question, and workers are paid
\$0.10 on successful completion. When collecting preference judgments,
we pack eight document pairs and a test question into each page, and
pay workers \$0.15 on successful completion. The rationale behind the
different pays is that workers receive the same amount of \$0.0083 per
document read. Each row is shown to workers until three trusted
judgments have been collected.

\textbf{Selection of queries and documents.} Queries and documents are
sampled from the \textsc{Trec} Web Track 2013 \& 2014. From the 100
available queries, we sample a subset of twelve queries.\footnote{Queries are
available in \url{http://trec.nist.gov/data/webmain.html}.}
Among the
sampled queries, one query is marked as ambiguous by
\textsc{Trec}, five queries
are marked as unambiguous (single), and six queries are faceted. The
original relevance judgments contain up to six relevance levels: junk
pages (\textit{Junk}), non-relevant (\textit{NRel}), relevant
(\textit{Rel}), highly relevant (\textit{HRel}), key pages
(\textit{Key}), and navigational pages (\textit{Nav}), corresponding
to six graded levels, i.e., -2, 0, 1, 2, 3, 4.  
Different from other grades, \textit{Nav} indicates a document
can satisfy a navigational user intend, making the comparison
relative to other documents depend on the information intent from
the crowdsourcing judges.
Hence, in our work, documents labeled \textit{Nav}
together with documents labeled \textit{Junk} are removed.  Due to the
limit occurrences, documents labeled \textit{Key} and \textit{HRel}
are both regarded as highly relevant.  For each query we determine two
sets of documents. Each set consists of twelve documents selected
uniformly across graded levels, resulting in four documents per
graded level. The first set is used to collect judgments; the second
set serves to create test questions. When collecting graded judgments,
the selected documents are directly used. To collect preference
judgments, we generate for each query all 66 pairs of documents and
randomly permute each document pair. Test
questions are generated treating the judgments from \textsc{Trec} as
ground truth. To ensure that workers on CrowdFlower see the same
documents as trained judges from \textsc{Trec}, we host copies of
ClueWeb12\footnote{\url{http://lemurproject.org/clueweb12/index.php}}
documents on our own web server.

\textbf{Time consumption.} To monitor the time consumed for reading
documents and making judgments, we proceed as follows. We record the
timestamp when judges start reading the shown document(s). To display
available options for judging, workers have to click on a button
``Click here to judge'', and we record the instant when this
happens. As a last timestamp, we record when the worker selects the
submitted option. 
In recording timestamps, the order of clicks from judges are restricted
by customized JavaScript, e.g., ``Click here to judge'' button is enabled
only after document(s) is (are) read.
We thus end up with three timestamps, allowing us to
estimate the reading time, as the time passed between the first two
timestamps, and the judgment time, as the time passed between the last
two.

\textbf{Judgment aggregation.} As mentioned, at least three trusted
judgments are collected for each row. One straightforward option to
aggregate them is to use majority voting as suggested by Alonso and
Mizzaro~\cite{alonso2011design}. However, in our setting, a simple
majority vote may not break ties, given that there are more than two
options to choose from. As a remedy we use workers' accuracies, as
measured on test questions, in a weighted majority voting to break ties.


\section{Results} 
\label{sec.result}

We now report the results of our empirical study. After giving some
general statistics about the collected judgments, we answer our three
research questions, by comparing
different groups of judgments over the same set of test queries
employing statistical instruments like Student's t-test. 
\bigbreak
\subsection{General Statistics}

Table~\ref{tab.descriptstat} summarizes general statistics about the
collected judgments. 
The collected judgments are publicly 
available.\footnote{\url{http://people.mpi-inf.mpg.de/~khui/data/ecir17empirical}}

\begin{table}[!t]
\centering

\caption{General statistics about judgments collected using crowdsourcing.}\label{tab.descriptstat}

\noindent\begin{tabularx}{\linewidth}{|X|X|X|X|}
\hline
\hfill\hfill\null& \hfill Graded Judgments\hfill\null
&\hfill Strict Preferences\hfill\null&\hfill Weak Preferences\hfill\null\\
\hline
\hline
\hfill Total Cost \hfill\null&\hfill$\$9.36$\hfill\null&\hfill$\$62.10$\hfill\null&$\hfill\$76.80$\hfill\null\\
\hline
\hfill\#Judgments\hfill\null&\hfill919\hfill\null&\hfill2,760\hfill\null&\hfill2,931\hfill\null\\
\hline
\hfill\#Judgments per Judge\hfill\null&\hfill28.80\hfill\null&\hfill55.00\hfill\null&\hfill20.10\hfill\null\\
\hline
\hfill Fleiss' $\kappa$\hfill\null&\hfill$0.170$\hfill\null&\hfill$0.498$\hfill\null&\hfill$0.253$\hfill\null\\
\hline
\hline
\multicolumn{4}{|c|}{\textit{Distribution of Judgments}}\\
\end{tabularx}\offinterlineskip
\\
\noindent\begin{tabularx}{\linewidth}{|X|X|X|}  \hline
\hfill``Highly-Relevant''\hfill\hfill28\% \hfill\null&\hfill$A\succ B$\hfill51\%\hfill\null&\hfill$A\succ B$\hfill30\%\hfill\null\\\hline
\hfill``Relevant''\hfill\hfill\hfill\hfill\hfill43\% \hfill\null&\hfill$A\prec B$\hfill49\%\hfill\null&\hfill$A\prec B$\hfill31\%\hfill\null\\\hline
\hfill``Non-Relevant''\hfill\hfill\hfill29\% \hfill\null&\hfill-\hfill\null&\hfill$A\sim B$\hfill39\%\hfill\null\\
\hline
\end{tabularx}
\end{table}

\textbf{Inter-judge agreement.}  Similar to~\cite{alonso2012using},
Fleiss' $\kappa$ is computed over each query and average Fleiss'
$\kappa$ among all queries is reported in
Table~\ref{tab.descriptstat}.  To put our results in context, we merge
``Highly-Relevant'' with ``Relevant'' and convert graded to binary
judgments, ending up with Fleiss' $\kappa=0.269$, which is close to
$0.195$ reported in~\cite{alonso2012using}.
In addition,
Kazai et al.~\cite{kazai2013user} reported  Fleiss' $\kappa=0.24$ (cf. Table 2 PC (e) therein)
among weak preference judgments from crowdsourcing, 
which approximates $0.253$ in our work.
We further conduct two-tailed Student's t-test in between the
three kinds of judgments over different queries. The p-value between
strict preferences and graded judgments is smaller than $0.001$;
between weak preferences and graded judgments is
$0.314$; whereas it is $0.005$ between the two kinds of
preference judgments. It can be seen that the judges achieve
better inter-agreement for strict preferences than for the
others, meanwhile there is no significant difference between
weak preferences and graded judgments.  This aligns
with the observations from~\cite{carterette2008here}, that
strict preferences exhibit higher inter-judges
agreement. The introduction of ``ties'' reduces the inter-judges
agreement, which might due to more options are available.

\begin{table*}[!t]
\centering
\caption{Transitivity over aggregated judgments. The ratio of transitive triples
  out of triples in different types is reported. The numbers in the bracket are the number of 
  transitive triples divides the total number of triples.}\label{tab.tranagg}
\resizebox{0.8\columnwidth}{!}{
\begin{tabular}{|c|c|c|c|c|c|}
\hline
\multirow{2}{*}{Query}&Strict Preferences&\multicolumn{4}{c|}{Weak Preferences}\\
\cline{2-6}
&\textit{asymTran}&\textit{asymTran}&\textit{s2aTran}&\textit{s2sTran}&Overall\\
\hline
\hline
216&100\% ($\nicefrac{220}{220}$)&96\% ($\nicefrac{78}{81}$)&89\% ($\nicefrac{90}{101}$)&8\% ($\nicefrac{3}{38}$)&78\% ($\nicefrac{171}{220}$)\\
\hline
222&99\% ($\nicefrac{218}{220}$)&100\% ($\nicefrac{40}{40}$)&98\% ($\nicefrac{117}{120}$)&50\% ($\nicefrac{30}{60}$)&85\% ($\nicefrac{187}{220}$)\\
\hline
226& 96\% ($\nicefrac{210}{220}$)&98\% ($\nicefrac{39}{40}$)&87\% ($\nicefrac{86}{99}$)&24\% ($\nicefrac{19}{81}$)&66\% ($\nicefrac{144}{220}$)\\
\hline
231& 98\% ($\nicefrac{216}{220}$)&100\% ($\nicefrac{17}{17}$)&95\% ($\nicefrac{107}{113}$)&30\% ($\nicefrac{27}{90}$)&69\% ($\nicefrac{151}{220}$)\\
\hline
241& 99\% ($\nicefrac{217}{220}$)&100\% ($\nicefrac{52}{52}$)&99\% ($\nicefrac{112}{113}$)&31\% ($\nicefrac{17}{55}$)&82\% ($\nicefrac{181}{220}$)\\
\hline
253& 91\% ($\nicefrac{199}{220}$)&100\% ($\nicefrac{24}{24}$)&86\% ($\nicefrac{66}{77}$)&38\% ($\nicefrac{45}{119}$)&61\% ($\nicefrac{135}{220}$)\\
\hline
254&99\% ($\nicefrac{218}{220}$)&100\% ($\nicefrac{39}{39}$)&97\% ($\nicefrac{105}{108}$)&36\% ($\nicefrac{26}{73}$)&77\% ($\nicefrac{170}{220}$)\\
\hline
257& 95\% ($\nicefrac{208}{220}$)&97\% ($\nicefrac{88}{91}$)&86\% ($\nicefrac{87}{101}$)&11\% ($\nicefrac{3}{28}$)&81\% ($\nicefrac{178}{220}$)\\
\hline
266&94\% ($\nicefrac{207}{220}$)&100\% ($\nicefrac{69}{69}$)&98\% ($\nicefrac{123}{125}$)&50\% ($\nicefrac{13}{26}$)&93\% ($\nicefrac{205}{220}$)\\
\hline
277&91\% ($\nicefrac{200}{220}$)&100\% ($\nicefrac{37}{37}$)&82\% ($\nicefrac{109}{133}$)&54\% ($\nicefrac{27}{50}$)&79\% ($\nicefrac{173}{220}$)\\
\hline
280&99\% ($\nicefrac{218}{220}$)&100\% ($\nicefrac{37}{37}$)&85\% ($\nicefrac{85}{100}$)&29\% ($\nicefrac{24}{83}$)&66\% ($\nicefrac{146}{220}$)\\
\hline
296&96\% ($\nicefrac{212}{220}$)&90\% ($\nicefrac{35}{39}$)&77\% ($\nicefrac{82}{106}$)&19\% ($\nicefrac{14}{75}$)&60\% ($\nicefrac{131}{220}$)\\
\hline
\hline
Avg.&96\% ($\nicefrac{212}{220}$)&98\% ($\nicefrac{46}{47}$)&90\% ($\nicefrac{98}{108}$)&32\% ($\nicefrac{21}{65}$)&75\% ($\nicefrac{164}{220}$)\\
\hline
\end{tabular}

}
\end{table*}

\subsection{RQ1: Transitivity}

In this section, transitivity is examined over both strict
and weak preference judgments. 
Different from in~\cite{carterette2008here} and in~\cite{rorvig1990simple},
we investigate transitivity based on
aggregated judgments. 
This is because the aggregated judgments are the
ultimate outcome from crowdsourcing, and also because, as mentioned in
Section~\ref{sec.introduction}, triples from a single judge are too
few over individual queries to lead to any conclusions. The results
per query are summarized in Table~\ref{tab.tranagg}. It can be seen
that over strict preferences, transitivity holds for 96\%
triples on average, and the number is between 91\% and 100\% over
individual query. This number is close to the transitivity reported
in~\cite{carterette2008here}, where average transitivity is
99\% and at least 98\% triples from a single judge are transitive.
Meanwhile, for weak preferences, this number is only 75\% on
average, and the minimum percentage is
60\% from query 296, indicating that transitivity does not hold in
general. To explore the reasons, we further decompose transitivity
according to different types of preferences within unique document
triples. In particular, the ``better than'' and ``worse than'' options are
referred as asymmetric relationships and the ``tie'' option is referred
as symmetric relationship~\cite{sep-preferences}. The transitivity can
be categorized as: \textit{asymTran}, which lies among asymmetric
relationships (no tie judgment in a triple); 
\textit{s2aTran}, which lies in between
symmetric and asymmetric relationships (only one tie judgment in a triple)
and \textit{s2sTran}, which lies among 
symmetric relationships  (at least two tie judgments in a triple).
 Over each query, the $220$
triples are thereby categorized according to the three types on which
transitive percentage is computed. From Table~\ref{tab.tranagg}, we
can see that \textit{asymTran} holds even better than in
strict preferences, meanwhile, \textit{s2aTran} holds for
$90\%$ on average. However, over \textit{s2sTran}, the transitivity
does not hold anymore: the transitive percentage drops to 32\% on
average.

\textbf{Answer to RQ1:} We conclude that transitivity holds for over 90\% 
aggregated strict preference judgments.
For weak preference judgments, though, transitivity only
holds among non-tie judgments (\textit{asymTran}) and in between tie
and non-tie judgments (\textit{s2aTran}). Thus, given judgments
$d_1\sim d_2$ and $d_2 \sim d_3$, we can not infer $d_1\sim d_3$. 
We can
see that, in terms of transitivity, weak and
strict preference judgments exhibit differently, and extra caution
must be taken when assuming transitivity when collecting weak
  preferences via crowdsourcing.

\begin{table}[!t]
\centering
\caption{Average time consumption (in seconds) and quartiles over twelve queries.}\label{tab.time}
\resizebox{0.75\columnwidth}{!}{
\begin{tabular}{|c|c|c|c|c|c|}
\hline
\multicolumn{2}{|c|}{Time Consumption}& average & $25^{th}$ percentile&median&$75^{th}$ percentile\\
\hline
\hline
\multirow{2}{*}{graded judgments}& Judgment&2.60&1.37&1.52&1.82\\
\cline{2-6}
& Total&24.24&11.73&19.55&28.88\\
\hline
\multirow{2}{*}{strict preferences}& Judgment&1.79&1.24&1.37&1.58\\
\cline{2-6}
& Total&34.17&17.84&25.28&40.98\\
\hline
\multirow{2}{*}{weak preferences}& Judgment&2.07&1.40&1.57&1.91\\
\cline{2-6}
& Total&32.43&15.77&24.57&39.10\\
\hline
\end{tabular}
}
\end{table}
\subsection{RQ2: Time Consumption}

We compare time consumption for different kinds of judgments looking
both at total time, which includes the time for reading document(s)
and judgment time. The results are summarized in Table~\ref{tab.time},
based on aggregated statistics from twelve queries. For judgment time, it
can be seen that judges spend least time with strict
  preferences. The p-values from two-tailed Student's t-tests between
the three kinds of judgments are as follows. P-value equals 0.055
between strict preferences and graded judgments,
equals 0.196, between weak preferences and graded
  judgments, and equals 0.100 between the two kinds of
preference judgments. We can conclude that judges are
slightly but noticeably
faster in making judgments with strict
  preferences than in making the other two kinds of judgments,
meanwhile the difference between the time consumption with
weak preferences and with graded judgments is
insignificant. As for total time, Table~\ref{tab.time} demonstrates
that judges are significantly faster in finishing single
graded judgments after considering reading time, with
p-value from two-tailed Student's t-test is less than $0.001$ relative to both
preference judgments. However, there is no significant
difference for judges with weak and strict
  preferences -- the corresponding p-value equals 0.168.

\textbf{Answer to RQ2:} 
Judges are faster in making strict
  preference judgments. When considering total time, judges
need to read two documents in preference judgments, making
total time consumption higher. Moreover, when comparing
the two kinds of preference judgments, judges take
significantly less time with strict preferences, 
meanwhile there is no difference in terms of total time consumption. 
Compared with~\cite{carterette2008here,rorvig1990simple}, time consumption
is measured among judges from crowdsourcing, who are with more diverse reading and 
judging ability and might be less skillful than trained judges.
Actually, the web pages being judged require more than 20 seconds on average to read, making reading
time dominate the total time consumption. 

\begin{table}[!t]
\centering
\caption{Agreement between graded judgments from crowdsourcing (columns) and \textsc{Trec} (rows).}
\resizebox{0.6\columnwidth}{!}{
\label{tab.grad2trec}
\begin{tabular}{|c|c|c|c||c|}
\hline
\backslashbox{{\scriptsize \textsc{Trec}}}{}
&\textit{Non-Relevant}& \textit{Relevant} &\textit{Highly-Relevant}&\#Total\\
\hline
\hline
\textit{Non-Relevant}&56.3\%&39.6\%&4.1\%&48\\
\textit{Relevant}&14.6\%&54.2\%&31.2\%&48\\
\textit{Highly-Relevant}&14.6\%&37.5\%&47.9\%&48\\
\hline
\end{tabular}
}
\end{table}

\begin{table}[!t]
\centering
\caption{Agreement between preference judgments from 
crowdsourcing (columns) and the one inferred from 
\textsc{Trec} judgments (rows).}\label{tab.pref2trec}
\resizebox{0.85\columnwidth}{!}{
\parbox{.49\linewidth}{
\centering

\begin{tabular}{|c|c|c||c|}
\multicolumn{4}{c}{(a) strict preferences}\\
\hline
\backslashbox{{\scriptsize \textsc{Trec}}}{}
&$A\prec B$& $A\succ B$&\#Total\\
\hline
\hline
$A\prec B$&83.0\%&17.0\%&282\\
$A\sim B$&46.8\%&53.2\%&216\\
$A\succ B$&20.4\%&79.6\%&294\\
\hline
\end{tabular}

}
\hfill
\parbox{.49\linewidth}{
\centering
\begin{tabular}{|c|c|c|c||c|}
\multicolumn{5}{c}{(b) weak preferences}\\
\hline
\backslashbox{{\scriptsize \textsc{Trec}}}{}
&$A\prec B$& $A\sim B$ &$A\succ B$&\#Total\\
\hline
\hline
$A\prec B$&62.8\%&30.9\%&6.3\%&285\\
$A\sim B$&17.6\%&59.7\%&22.7\%&216\\
$A\succ B$&7.6\%&32.0\%&60.5\%&291\\
\hline
\end{tabular}

}
}
\end{table}

\subsection{RQ3: Quality}

We compare the quality of three kinds of judgment
collected via crowdsourcing
in terms of their agreement with judgments from \textsc{Trec} (qrel). 
We employ both percentage agreement, which counts the
agreed judgments and divides it by the number of total judgments, and
Cohen's $\kappa$ as in~\cite{alonso2012using}, and use the latter 
for two-tailed Student's t-tests. 
When evaluating preference judgments from crowdsourcing, judgments from
\textsc{Trec}  are first converted to
preference judgments, by comparing labels over two
documents, resulting in ``better than'', ``worse than'' or ``tie''.
The percentage agreement over three kinds of
judgment relative to judgments from \textsc{Trec} are summarized in
Table~\ref{tab.grad2trec} and~\ref{tab.pref2trec}, where the percentage
is normalized per row.
To put our results in context, we first measure agreement based on
binary judgments, by merging the grades \textit{Relevant} and
\textit{High-Relevant} in both \textsc{Trec} judgments and
graded judgments from crowdsourcing.
In~\cite{alonso2012using}, percentage agreement equals $77\%$ and
Cohen's $\kappa=0.478$, relative to judgments from TREC-7
and TREC-8.  
Meanwhile we obtain $75.7\%$ and Cohen's
$\kappa=0.43$ -- slightly lower values. We argue that is due to the
document collections in use: ClueWeb12, used in our work, consists
of web pages which are more diverse and noisy, making it harder to
judge; whereas disk 4\&5 used in TREC-7 and TREC-8 consist of cleaner
articles.\footnote{\url{http://trec.nist.gov/data/docs_eng.html}}
When using three grades,
graded judgments from crowdsourcing
achieve $52.8\%$ and
Cohen's $\kappa=0.292$ relative to judgments from \textsc{Trec}. 
And the percentage agreement is $59.1\%$ and Cohen's $\kappa=0.358$ for
strict preferences, whereas for weak preferences the
numbers are $61\%$ and $0.419$ respectively. 
Compared with graded judgments from crowdsourcing,
the corresponding p-values from paired sample t-tests
over Cohen's $\kappa$ among queries are 0.259 and 0.052, 
indicating weak preference judgments agree with
\textsc{Trec} judgments better.

Note that, however, for documents with the same grade in
\textsc{Trec} a tie is inferred, whereas strict preferences
do not permit tie judgments. From Table~\ref{tab.pref2trec}~(a), it
can be seen that 216 document pairs are inferred as tied, 
where agreement is zero for strict
  preferences currently. To mitigate this mismatch, in line 
  with~\cite{carterette2008here}, tie judgments in inferred
preference judgments are redistributed as ``A is better'' or
``B is better''. In this redistribution, an agreement is assumed, 
coined as $aar$.  In other
words, the 216 document pairs that are inferred as tied
in~Table~\ref{tab.pref2trec}~(a) are redistributed so that
$216\times aar$ random pairs are assigned with the same judgments as in
collected strict preference judgments. The logic behind this is that the
ground-truth strict preferences over these inferred ties are
unknown and we need to assume an agreement over them to make
strict preference judgments comparable.
Thereby,
two groups of agreement are reported for strict preference judgments 
at assumed agreement
rates $aar=50\%$ and $80\%$, respectively corresponding to random agreement and
the average agreement under non-tie situations (average of $83\%$ and $79.6\%$ in
Table~\ref{tab.pref2trec}~(a)). 
Without influencing comparison results,
graded judgments from crowdsourcing are also converted
to preference judgments, making three kinds
of judgments from crowdsourcing more comparable.
In Table~\ref{tab.kappa}, it can be seen
that Cohen's $\kappa=0.530$ for strict preferences when assuming 
$aar=50\%$, and the value for weak preferences is $0.419$. 
Both preference judgments
agree with \textsc{Trec} significantly better than graded
  judgments, with p-values from paired sample t-test equal $0.001$
and $0.015$ respectively.  We further compare Cohen's $\kappa$ from
strict preferences ($aar=50\%$) with the one from weak
  preferences, getting p-value from paired sample t-test equals
$0.004$, indicating strict preference judgments agree with judgments from
\textsc{Trec} significantly better than weak preferences.

\begin{table}[!t]
\centering
\caption{Percentage agreement and 
  Cohen's $\kappa$ between inferred preference judgments
  from \textsc{Trec} and three kinds 
  of judgments collected via crowdsourcing. 
  For the column of strict preferences,
  tie judgments in the inferred judgments from 
  \textsc{Trec} are redistributed by assuming
  different agreement rates. Results under 
  $aar=50\%$ and $80\%$ are reported.}\label{tab.kappa}
\resizebox{0.8\columnwidth}{!}{
\begin{tabular}{|c|c|c|c|c|c|c|c|c|c|}
\hline
\multirow{3}{*}{Query}
&\multicolumn{4}{c}{Strict Preferences} & 
\multicolumn{2}{|c|}{\multirow{2}{*}{Weak Preferences}}&
\multicolumn{2}{c|}{\multirow{2}{*}{Graded Judgments}}\\
\cline{2-5}
&\multicolumn{2}{c|}{break tie $aar=50\%$}&
\multicolumn{2}{c|}{break tie $aar=80\%$}&\multicolumn{2}{c|}{}&\multicolumn{2}{c|}{}\\
\cline{2-9}
&{\scriptsize percentage}&{\scriptsize Cohen's $\kappa$}
&{\scriptsize percentage}&{\scriptsize Cohen's $\kappa$}
&{\scriptsize percentage}&{\scriptsize Cohen's $\kappa$}
&{\scriptsize percentage}&{\scriptsize Cohen's $\kappa$}\\
\hline
\hline
216&77\%&0.594&85\%&0.710&65\%&0.466&53\%&0.269\\
222&76\%&0.569&83\%&0.680&59\%&0.391&65\%&0.474\\
226&77\%&0.589&79\%&0.611&65\%&0.473&62\%&0.386\\
231&70\%&0.494&83\%&0.686&53\%&0.310&65\%&0.435\\
241&74\%&0.557&83\%&0.689&70\%&0.543&59\%&0.386\\
253&74\%&0.533&77\%&0.576&49\%&0.248&36\%&0.044\\
254&80\%&0.649&91\%&0.821&71\%&0.573&65\%&0.471\\
257&73\%&0.529&83\%&0.680&64\%&0.445&61\%&0.380\\
266&70\%&0.459&73\%&0.500&73\%&0.588&38\%&0.048\\
277&68\%&0.397&70\%&0.417&50\%&0.261&38\%&0.075\\
280&65\%&0.389&74\%&0.510&56\%&0.345&44\%&0.193\\
296&77\%&0.601&85\%&0.715&59\%&0.386&50\%&0.224\\
\hline
Avg&74\%&0.530&81\%&0.633&61\%&0.419&53\%&0.282\\
\hline
\end{tabular}
}
\end{table}

\textbf{Answer to RQ3:} From Table~\ref{tab.kappa}, it can be seen
that agreement from strict preferences under $aar=50\%$ and weak
  preferences are $88\%$ and $49\%$ higher than the collected graded
  judgments in terms of Cohen's $\kappa$. 
  We further compare this
agreement relative to \textsc{Trec} 
with the agreement among trained
judges reported in literature, similar to~\cite{alonso2012using}.
Intuitively, if agreement between judgments from crowdsourcing and
from \textsc{Trec} is comparable to the one among
trained judges, we can conclude that judgments from crowdsourcing are
good enough to replace those from trained judges.
Carterette et al.~\cite{carterette2008here} reported agreement among six
trained judges over preference judgments, 
and the percentage agreement is
$74.5\%$ (cf. Table~2~(a) therein), whereas 
in our work agreement for strict
  preferences are $74\%$ under $aar=50\%$, and $81\%$ under $aar=80\%$.  
Kazai et al.~\cite{kazai2013user} reported that
Fleiss' $\kappa$ among trained judges over preference
  judgments is $0.54$ (cf. Table 2 PE (e) therein). Thus, we recompute
  the agreement between strict preference judgments and judgments from \textsc{Trec}
  in terms of 
Fleiss' $\kappa$, and get
$\kappa=0.504$ under $aar=50\%$ and $0.637$ under $aar=80\%$. Note that
strict preferences are collected in~\cite{carterette2008here}
and weak preferences are employed in~\cite{kazai2013user}. Since
the difference of these two kinds of preference judgments
when collected from trained judges is unclear, we regard them the same.
We can conclude that the agreement between strict
  preferences collected via crowdsourcing and \textsc{Trec} are
comparable to the one among trained judges. 
Moreover,
compared with strict preference judgments,
we can conclude that judgment
quality in crowdsourcing
is significantly degraded when using 
weak preferences.
 
As reported in~\cite{alonso2009can,alonso2012using}, we also observe
judges from crowdsourcing can sometimes point out mistakes 
in \textsc{Trec} judgments. 
In total, we found around 20 such documents, especially
via ``test questions'', by examining documents (or document pairs)
that receive majority judgments opposing to the judgments from
\textsc{Trec}. 
One example is\\
clueweb12-0013wb-31-22050
and
clueweb12-0806wb-32-26209
for query 280, ``view my internet history''.  The former is labeled as
``Highly-Relevant'' and the latter is labeled as ``Relevant'' in qrel.
However, none of them is relevant: the first page is a comprehensive
list about history of internet \& W3C, and the second page is a question
on a forum about how to clean part of ones' browsing history.
\bigskip

\subsection{Discussion}
It has been demonstrated that
weak and strict preferences are different in all three regards.
To investigate the reasons,
we reduce the number of options in weak preferences by merging
``tie'' with ``A is better'', merging ``tie'' with ``B is better'' or
merging the two non-tie options, measuring the agreements among judges, 
getting Fleiss' $\kappa=0.247$, $0.266$, and $0.073$ respectively.  
The corresponding p-values from
two-tailed Student's t-tests relative to the one with three options are $0.913$,
$0.718$, and less than $0.001$. 
It can be seen that judges tend to disagree more when making choices between
ties and non-ties judgments.
Put differently, 
the threshold to make a non-tie judgment
is ambiguous and is varied among different judges.
This implies that the tie option
actually makes the judgments more complicated, namely,
judges have to firstly determine whether the difference is large
enough to be non-tied before judging the preferences.


\section{Conclusion} 
\label{sec.conclusion}

In this work, we use crowdsourcing to collect graded
  judgments and two kinds of preference judgments.  
In terms of judgment quality, the three kinds of judgments can be
sorted as follows, graded judgments $<$ weak $<$
strict preference judgments.  
  Moreover, our position for tie judgments is: it can be used but must be
with more cautions when collected via crowdsourcing, especially when
attempting to assume transitivity.


\bibliographystyle{splncs03}

\begin{thebibliography}{10}
\providecommand{\url}[1]{\texttt{#1}}
\providecommand{\urlprefix}{URL }

\bibitem{alonso2011design}
Alonso, O., Baeza-Yates, R.: Design and implementation of relevance assessments
  using crowdsourcing. {\em ECIR}~2011.

\bibitem{alonso2009can}
Alonso, O., Mizzaro, S.: Can we get rid of trec assessors? Using mechanical
  turk for relevance assessment. {\em SIGIR} 2009 Workshop on
  the Future of IR Evaluation.

\bibitem{alonso2012using}
Alonso, O., Mizzaro, S.: Using crowdsourcing for trec relevance assessment.
  Information Processing \& Management 48(6), 2012.

\bibitem{bashir2013document}
Bashir, M., Anderton, J., Wu, J., Golbus, P.B., Pavlu, V., Aslam, J.A.: A
  document rating system for preference judgements. {\em SIGIR} 2013.

\bibitem{carterette2008here}
Carterette, B., Bennett, P.N., Chickering, D.M., Dumais, S.T.: Here or there:
  {Preference Judgments for Relevance}. {\em ECIR} 2008.

\bibitem{cleverdon1967cranfield}
Cleverdon, C.: The cranfield tests on index language devices. Aslib
  proceedings. vol.~19. 1967.

\bibitem{grady2010crowdsourcing}
Grady, C., Lease, M.: Crowdsourcing document relevance assessment with
  mechanical turk. {\em NAACL HLT} 2010 workshop on creating
  speech and language data with Amazon's mechanical turk.

\bibitem{sep-preferences}
Hansson, S.O., Grüne-Yanoff, T.: Preferences. Zalta, E.N. (ed.) The
  Stanford Encyclopedia of Philosophy. 2012.

\bibitem{kazai2011search}
Kazai, G.: In search of quality in crowdsourcing for search engine evaluation.
  {\em ECIR} 2011.
  
\bibitem{kazai2013user}
Kazai, G., Yilmaz, E., Craswell, N., Tahaghoghi, S.M.: User intent and assessor
  disagreement in web search evaluation. {\em CIKM} 2013.

\bibitem{moshfeghi2016identify}
Moshfeghi, Y., Huertas-Rosero, A.F., Jose, J.M.: Identifying careless workers
  in crowdsourcing platforms: A game theory approach. {\em SIGIR} 2016.

\bibitem{moshfeghi2016game}
Moshfeghi, Y., Rosero, A.F.H., Jose, J.M.: A game-theory approach for effective
  crowdsource-based relevance assessment. ACM Trans. Intell. Syst. Technol.
  7(4), 2016.

\bibitem{Radinsky2011}
Radinsky, K., Ailon, N.: Ranking from pairs and triplets: Information quality,
  evaluation methods and query complexity. {\em WSDM} 2011.

\bibitem{rorvig1990simple}
Rorvig, M.E.: The simple scalability of documents. Journal of the American
  Society for Information Science  41(8), 1990.

\bibitem{song2011select}
Song, R., Guo, Q., Zhang, R., Xin, G., Wen, J.R., Yu, Y., Hon, H.W.:
  Select-the-best-ones: A new way to judge relative relevance. Information
  Processing \& Management  47(1), 2011.

\bibitem{zhu2010analysis}
Zhu, D., Carterette, B.: An analysis of assessor behavior in crowdsourced
  preference judgments. {\em SIGIR} 2010 workshop on crowdsourcing for search
  evaluation.

\end{thebibliography}

\end{document}